%% file: main.tex
\documentclass[sigconf,authorversion]{acmart}

\usepackage{libertine}
\usepackage[T1]{fontenc}
\usepackage[utf8]{inputenc}
\usepackage{hyperref}
\usepackage{natbib}
\usepackage[symbol]{footmisc}
\renewcommand{\thefootnote}{\fnsymbol{footnote}}

\acmYear{2020}\copyrightyear{2020}
\setcopyright{acmcopyright}
\acmConference[FAT* '20]{Conference on Fairness, Accountability, and Transparency}{January 27--30, 2020}{Barcelona, Spain}
\acmBooktitle{Conference on Fairness, Accountability, and Transparency (FAT* '20), January 27--30, 2020, Barcelona, Spain}
\acmPrice{15.00}
\acmDOI{10.1145/3351095.3372853}
\acmISBN{978-1-4503-6936-7/20/01}

\renewcommand\footnotetextcopyrightpermission[1]{} 
\input{parts/packages}
\input{parts/macros}

\title{POTs: Protective Optimization Technologies}

\author{Bogdan Kulynych}
\affiliation{EPFL}
\email{bogdan.kulynych@epfl.ch}

\author{Rebekah Overdorf}
\affiliation{EPFL}
\email{rebekah.overdorf@epfl.ch}

\author{Carmela Troncoso}
\affiliation{EPFL}
\email{carmela.troncoso@epfl.ch}

\author{Seda Gürses}
\affiliation{TU Delft / KU Leuven}
\email{f.s.gurses@tudelft.nl}

\begin{abstract}
Algorithmic fairness aims to address the economic, moral, social, and political impact that digital
systems have on populations through solutions that can be applied by service providers.
Fairness frameworks do so, in part, by mapping these problems to a narrow definition
and assuming the service providers can be trusted to deploy countermeasures.  Not
surprisingly, these decisions limit fairness frameworks' ability to capture a
variety of harms caused by systems.

We characterize fairness limitations using
concepts from requirements engineering and from social sciences. We show that the focus on
algorithms' inputs and outputs misses harms that arise from systems interacting with the world;
that the focus on bias and discrimination omits broader harms on populations and their
environments; and that relying on service providers excludes scenarios where they are
not cooperative or intentionally adversarial.

We propose \emph{Protective Optimization Technologies (POTs)}.
POTs, provide means for affected parties to address the negative impacts of systems in the
environment, expanding avenues for political contestation. POTs intervene from outside the system,
do not require service providers to cooperate, and can serve to correct, shift, or expose harms that systems impose on populations
and their environments. We illustrate the potential and limitations of POTs in two case studies:
countering road congestion caused by traffic-beating applications, and recalibrating
credit scoring for loan applicants.
\end{abstract}

\begin{document}

\maketitle
\blfootnote{Bogdan Kulynych and Rebekah Overdorf contributed equally to this work.}

\input{parts/intro}
\input{parts/jackson}
\input{parts/fairness_critique}
\input{parts/pots}

\input{parts/eval}

\section{Discussion}
\input{parts/discussion}

\section*{Acknowledgements}
We are indebted to Martha Poon for her original framing of the optimization problem. We also thank
Ero Balsa for his collaboration on the previous versions of this work, Jakub Tarnawski for his
invaluable help with finding the right tool for the traffic-routing case study, and Muhammad Bilal
Zafar for his helpful comments.

This work was supported in part by the Research Council KULeuven: C16/15/058, and the European
Commission through KULeuven BOF OT/13/070, H2020-DS-2014-653497 PANORAMIX, and H2020-ICT-2015-688722
NEXTLEAP. Seda Gürses is supported by a Research Foundation--Flanders (FWO) Fellowship. Rebekah
Overdorf is supported by the Open Technology Fund--Information Controls Fellowship.

\newpage

\bibliographystyle{unsrt}
\bibliography{main}

\clearpage

\appendix
\input{parts/appendices}

\end{document}

%% file: parts/packages.tex
\usepackage[utf8]{inputenc}
\usepackage{natbib}
\usepackage{url}
\usepackage{amsmath}
\usepackage{amssymb}
\usepackage{amsthm}
\usepackage{bm}
\usepackage{venndiagram}
\usepackage{dashbox}
\usepackage{xspace}
\usepackage{tikzsymbols}
\usepackage{dutchcal}
\usepackage{xcolor}
\usepackage{wasysym}
\usepackage{algorithm}
\usepackage{algorithmicx}
\usepackage{caption}
\usepackage{subcaption}
\usepackage{tikz}
\usepackage{graphicx}

%% file: parts/macros.tex
\newcommand{\seda}[1]{}
\newcommand{\bogdan}[1]{}
\newcommand{\bknote}[1]{}
\newcommand{\carmela}[1]{}
\newcommand{\bekah}[1]{}

\newcommand{\newterm}{\emph}
\newcommand{\Eqref}[1]{Eq.~\ref{#1}}
\newcommand{\Secref}[1]{Sec.~\ref{#1}}
\newcommand{\Figref}[1]{Fig.~\ref{#1}}
\newcommand{\parait}[1]{\smallskip\noindent\textit{#1}}

\newcommand\blfootnote[1]{%
  \begingroup
  \renewcommand\thefootnote{}\footnote{#1}%
  \addtocounter{footnote}{-1}%
  \endgroup
}

\newcommand{\define}{\triangleq}

\newcommand{\fair}{{\textit{fair}}}

\newcommand{\sR}{\mathbb{R}}

\newcommand{\sV}{\mathbb{V}}
\newcommand{\sE}{\mathbb{E}}
\newcommand{\gG}{\mathcal{G}}

\newcommand{\Pareto}{\Theta^*}
\newcommand{\sysutil}{U}
\newcommand{\benefit}{B}

\newcommand{\indbenefit}{B}

\newcommand{\objective}{J}
\newcommand{\correction}{\varepsilon}
\newcommand{\population}{P}

\newcommand{\syspar}{\mathbb{\theta}}
\newcommand{\hessian}{\mathcal{H}}
\newcommand{\influence}{IF_\varepsilon}

\newcommand{\misspecuser}{{i}}

\newcommand{\sysinputspace}{\mathcal{X}}

\newcommand{\sysinputvec}{\mathbf{x}}
\newcommand{\sysinput}{x}
\newcommand{\modcost}{C}
\newcommand{\opt}{\textit{opt}}
\renewcommand{\pot}{\textit{pot}}

\newcommand{\graph}{\gG}
\newcommand{\vertices}{\sV}
\newcommand{\edges}{\sE}
\newcommand{\lengthfunc}{s}
\newcommand{\speedlimfunc}{v}
\newcommand{\timefunc}{t}
\newcommand{\source}{a}
\newcommand{\sink}{b}
\newcommand{\timecost}{t}
\newcommand{\timethreshold}{t^*}
\newcommand{\graphpath}{\mathbf{e}}
\newcommand{\edgecost}{c}
\newcommand{\timedelta}{\Delta \timefunc}
\newcommand{\speeddelta}{\Delta \speedlimfunc}
\newcommand{\pathvar}{\pi}
\newcommand{\interdictvar}{\mu}

\newcommand{\poisonset}{X_\pot}
\newcommand{\pool}{X_\textit{pool}}
\newcommand{\reg}{R}
\newcommand{\collective}{G_\pot}

\newtheorem{theorem}{Theorem}[section]

\newtheorem{statement}{Statement}[section]
\newcommand{\Thmref}[1]{Theorem~\ref{#1}}

%% file: parts/intro.tex
\section{Introduction}
\label{sec:intro}

Advances in computational power, software engineering, and machine learning algorithms have been
instrumental in the rise of digital systems. Their ubiquity in our everyday
activities raises concerns regarding the centralization of decisional power~\cite{Yeung18}. These
concerns are amplified by the opaque and complex nature of these systems which results
in hard-to-explain outputs~\cite{MittelstadtRW19} and unjust outcomes for historically marginalized
populations~\cite{browne2015dark,gandy2016coming,AngwinLMK16,EnsignFNSV17}.

Computer scientists counter these inequities through frameworks studied under the
rubric of \emph{fairness}, using a variety of formal fairness notions~\cite{BarocasHN18}.
Their results have been instrumental in our understanding of the discriminatory effects of algorithmic decisions.
The frameworks, however, rely on narrowing the inequity problem to primarily consider
the discriminatory impact of algorithms, and assuming trustworthy providers.
The narrow view has enabled valuable breakthroughs centered around the service providers' ability to
address some inequities, but fails to capture broader harms or explore other ways to contest
service-provider power~\cite{OverdorfKBTG18}.

In this work we investigate digital systems from a new perspective in
order to understand how to address the broader class of harms that they cause.
To achieve this, we characterize the type of systems in which algorithms are integrated and deployed.
These systems typically build on distributed service architectures and
incorporate real-time feedback from both users and third-party service
providers~\cite{GursesVHoboken18, Kaldrack2015}. This feedback is leveraged for a variety of novel
forms of \emph{optimization} that are geared towards extraction of value through the system. Typically,
optimization is used for technical performance and minimizing costs, e.g., by optimizing cloud usage orchestration or
allocation of hardware resources. It has also become part and parcel of ``continuous development'' strategies based on experimentation
that allow developers to define dynamic objective functions and build adaptive systems.  Businesses
can now design for ``ideal'' interactions and environments by optimizing feature selection,
behavioral outcomes, and planning that is in line with a business growth strategy. We argue that
\emph{optimization-based systems} are developed to capture and manipulate behavior and environments for the extraction of value.
As such, they introduce broader risks and harms for users and environments beyond the outcome of a
single algorithm within that system. These impacts go beyond the bias and discrimination stemming from algorithmic outputs fairness frameworks consider.

Borrowing concepts and techniques from software and requirements engineering, and from economics and social sciences, we
characterize the limitations of current fairness approaches in capturing and mitigating harms
and risks arising from optimization. Among others, we show that focusing on the algorithms and their outputs overlooks the
many `externalities' caused by optimizing every aspect of a system; that discrimination is only one of the
injustices that can arise when systems are designed to maximize gain; and that ignoring service
providers' incentives and capabilities to enforce proposed solutions limits our understanding of
their operation and further consolidates the power providers have regarding decisions and
behaviors that have profound effects in society.

Finally, we propose \emph{Protective Optimization Technologies (POTs)}, which aim at addressing
risks and harms that cannot be captured from the fairness perspective and cannot be addressed
\emph{without} a cooperative service provider.  The ultimate goal of POTs is to eliminate the harms,
or at least mitigate them. When these are not possible, POTs can shift harms away from affected
users or expose abusive or non-social practices by service providers.  We illustrate the potential
of POTs to address externalities of optimization-based systems in two case studies: traffic-beating
routing applications and credit scoring. We also identify numerous techniques developed by
researchers in the fields of security and privacy, by artists, and by others, that, though not
necessarily designed to counter externalities, can be framed as POTs.




%% file: parts/jackson.tex
\section{Fairness from a system's perspective}
\label{sec:jackson}

We introduce Michael A. Jackson's theory of requirements engineering~\cite{Jackson95}
to discuss the focus, goals, and assumptions behind fairness framework from a systems' perspective.
This theory argues that computer scientists and engineers ``are concerned both with the
world, in which the \newterm{machine} serves a useful purpose, and with the machine itself [...] The
purpose of the machine [...] is located in the world in which the machine is to be installed and
used.'' In other words, our objective is to design systems that fulfill requirements \textit{in the}
\newterm{world} where the system is to be deployed.

More precisely, a portion of the world becomes the \textit{environment}, or the \textit{application
domain},\footnote{Application domain does not refer to a
class of applications, like health or banking domain, but to actors and things
in the environment where the machine is introduced.} of a \newterm{machine}. In this world there are
\textit{phenomena} in the application domain (e.g., events, behavior of people, activities) and in
the machine (e.g., data, algorithms, state). The machine's inputs and outputs, i.e., the things the
machine can sense or affect, are shared phenomena: they are observable both to the application domain and
the machine.


We introduce machines into existing application domains to effect change in these domains. To
achieve the desired change, we need descriptions of the phenomena before the machine is introduced,
known as \newterm{domain assumptions} $K$; and statements about the desired situation once the
machine is introduced to the domain, known as \newterm{requirements} $R$. A \newterm{specification}
$S$ is a set of requirements providing enough information for engineers to implement the machine.
$S$ typically describes phenomena shared between the machine and application domain. A
\newterm{program} $P$ derived from the specification is a description of a machine. If implemented
correctly, programs satisfy the specification. If the specification is derived correctly, programs
generate phenomena that attain the desired effects in the application domain, i.e., they
fulfill the requirements.

Jackson's explicit treatment of the
application domain and its interaction with the machine helps us to project known problems with
algorithms to a systems view. It enables us to distinguish problems due to badly derived
requirements (description of the problem) from those due to badly derived specifications
(description of the solution) and those due to badly implemented programs (how solutions are implemented)~\cite{jackson95book}.

\subsection{The focus on algorithms is insufficient for addressing inequitable outcomes of systems}
In the light of Jackson's theory, we claim that \textit{the focus on algorithms leaves out the systems' view.}
Algorithms are often a part of a larger technical system, which is deployed in an
environment. Fairness proposals rarely evaluate the systems' environmental conditions,
thereby leaving out the possibility that, \emph{even} when a fairness metric is satisfied by the
algorithm, the system (environment plus the machine) could still be unfair or have
other negative side effects. Such focus on the specification of the machine also promotes the
idea ``that the difficulty in addressing [unfairness] lies in devising a solution''~\cite{Jackson95}
and not necessarily in rethinking the world. 

Most fairness frameworks focus on describing
$S$ independent of $K$ and $R$ and then guarantee that states of
the machine, and its input and outputs, have certain properties. However, a machine that has
fair inputs and outputs and fulfills a specification $S_\fair$ does not guarantee the fulfillment of
requirements of fairness in the application domain. This would require evaluating $K$, 
establishing what phenomena the machine is expected to change in the application domain,
and articulating requirements $R_\fair$ with fairness as a goal.
Only then could we evaluate whether an $S_\fair$ fulfills $R_\fair$.

Focusing on $S_\fair$ has a number of repercussions. First, it does not reflect \emph{how harms
manifest themselves in the environment.} Without an understanding of $K$ and $R_\fair$,
a specification $S_\fair$ may simply not lead to a fair outcome.
Imagine a hypothetical ``fair predictive policing algorithm'' that can fairly
distribute police officers to different neighborhoods. If the algorithm does not
consider that the policing institution is already configured to control minorities
\cite{LavigneCT17} and that interactions with police pose greater risk of harms for minorities, a
``fair allocation'' can still disparately impact those minorities~\cite{EnsignFNSV17}.

Second, focusing on achieving fairness for users might \emph{leave out the impact of the system on
phenomena in the application domain that is not shared with the machine}. Unjust
outcomes could arise due to the optimization of certain behavior in the application domain,
and not because $S$ was unfair. For
instance, self-regulated housing markets such as Airbnb~\cite{airbnb_url} may not actively discriminate against their users, but
reports have shown that they can disrupt neighborhoods by changing rent dynamics and neighborhood
composition~\cite{airbnb-destroys}.

Third, \emph{the focus on algorithms abstracts away potential harms of phenomena in the machine.}
Much of the fairness literature focuses on ways in which algorithms can be biased or on harms caused
by decision-making algorithms. This overlooks that, when the system hosting the algorithms optimizes
its operation, 
it may gather more inputs and outputs than those of the algorithm.
Therefore, focusing on the algorithm may miss effects on the world that can go beyond those
generated by the outputs of the algorithm actions.

When phenomena in the machine domain are subject to optimization,
unfairness can arise from optimization programs $P_\opt$ fulfilling the specification
$S_\fair$ but not the requirements $R_\fair$ in the application domain.
For instance, prediction techniques to optimize targeted advertising can create discriminatory
effects~\cite{ali2019discrimination}; and exploration strategies to optimize $S_\fair$ may
gather inputs from the application domain that put some users unfairly at
risk~\cite{BirdBCW16}.

\subsection{Discriminatory effects are not the only concern for building just systems}
\label{sec:jackson-discrimination}

In Jackson's terms, considering only discriminatory effects constrains
the requirements $R_\fair$ to a particular class of harms. This
approach risks \emph{missing other harms caused by the system} when evaluating the performance of
the specification $S_\fair$ in the environment.

We assume the introduction of a machine in an environment aims to improve specific phenomena. 
The fact that this machine follows a specification $S_\fair$ that does not discriminate according to
$R_\fair$ does not guarantee that this machine will not induce other harms to the
environment in any applications domain. Take as examples unsafe housing or bad working conditions. 
If we use a machine to distribute these resources more efficiently, even if it does so fairly, 
\emph{all} users are harmed. Bad housing conditions and lack of labor protection 
are problems in and of themselves so users will be badly served regardless of fairness conditions.
When the system is by design unjust, or when the phenomena are structurally unjust or harmful,
claims on $S_\fair$ are meaningless.
In Jackson's terms: the requirements $R$ are incomplete with respect to just outcomes.


Sometimes, however, injustices are tightly tied to the machine: $S$ is not the solution but the
source of problems.  The way requirements $R$ are optimized might lead to externalities for (a
subpopulation of) users. When a specification $S$ optimizes an asocial outcome, such as excessive
user engagement in social networks~\cite{Tufekci18}, it can expose users to harms like addiction.
Solving fairness in this system will not resolve the underlying problem: the system is harmful.

Striving to fulfill the requirements $R_\fair$ \emph{itself} might bring new harms to the
application domain. Consider a fairness solution that alleviates distributional shift based on
increasing diversity in the training set. If its specification $S_\fair$ requires
collecting data from more individuals or collects new attributes to implement the fairness measure,
it will exacerbate privacy issues. These issues might result in many other harms in the
application domain.

This ontology assumes that $S$ is built to ``solve problems'' and ``improve phenomena'' in the world.
Thus, it does not provide the conceptual tools to address adverse situations.
This positive valence hinders the consideration of cases in which a machine amplifies existing
injustices or introduce new ones. Neither this ontology nor fairness frameworks account
for power imbalances or economic incentives, and how they impact how machine requirements are
considered and prioritized. We consider these matters in the next section by augmenting our
systems view to consider socio-economic aspects of the application domain.

%% file: parts/fairness_critique.tex
\section{Fairness, incentives, and power}
\label{sec:fairness-critique}

In this section, we extend our analysis to problems related to the political economy of systems.  To
model harms, we borrow the term \emph{negative externalities} from the economics literature.  A
system causes negative externalities when its consumption, production, and investment decisions
cause significant repercussions to users, non-users, or the environment~\cite{Starrett11}.  The
introduction of a machine might cause externalities in the application domain, independent of the
completeness or correctness of its requirements and specification.  For example, the heavy use of
traffic-beating apps such as Waze can worsen congestion for \emph{all} drivers in the application
domain~\cite{cabannes18}.  We argue that validating the specification $S$ against the requirements
$R$ is not enough. To build just systems one must consider externalities of the machine in the
application domain.

Congruent with models in fair optimization and economics, to express externalities and incentives
we introduce two utility functions that capture the machine's impact on the application
domain: the \newterm{service provider's utility}, which measures how much value the provider
extracts from introducing the machine, and the \newterm{social utility}, which measures the machine's utility
 for the environment and people. We define two versions of social utility: one
with a ``god's view'' of the application domain and one from the specification's perspective.
These utility functions enable us to capture \newterm{injustices} due to the introduction of the
machine into the environment.

\subsection{Idealized Fair-by-Design Service Provider}
We first consider an idealized \emph{fair-by-design service provider} that is willing to address the
externalities of the machine. That is, this provider aims to maximize both their own utility and the
social utility.  Using this setting, we show the ways in which fairness models fail to address a
broad class of systems' externalities.

We consider that a system is parametrized by a vector of internal parameters $\syspar \in \Theta$
for some convex set of possible parameters $\Theta$. Let $\population$ be a \emph{population}, a set
of individuals or other environmental entities that might be affected by the system. Let
$\sysutil(\syspar): \Theta \rightarrow \sR$ be the utility function of the provider when they
use parameters $\syspar$. Let $\indbenefit(\syspar): \Theta \rightarrow \sR$ denote a hypothetical
social-utility function, or \emph{benefit}, defined in the requirements $R$. Let $\hat
\benefit(\syspar)$ be the social utility in the provider's specification.
The provider optimizes its operation by solving the following multi-objective optimization problem:
\begin{equation}\label{eq:benevolent-opt}
    \max_{\syspar \in \Theta} \{ \sysutil(\syspar), \hat \benefit(\syspar) \}
\end{equation}
This problem is considered in fair learning literature in its scalarized form or
constraint-form~\cite{KamishimaAS11, ZafarVGG17a, ZafarVGG17b, CelisSV17, DoniniOBSP18}, with the
social utility modeling a notion of fairness. We assume
an ideal situation in which the chosen parameter vector $\syspar^*$ is a Pareto-optimal
solution~\cite{Miettinen12}, that is, it cannot improve any of the objectives without hurting at
least one of them. Pareto-optimality, however, is not sufficient to guarantee fairness or equity as different
trade-offs between the objectives are possible~\cite{LeGrand90}. We assume that out of the
possible Pareto solutions, the provider chooses one that maximizes social utility:
$
    \syspar^* \define \arg \max_{\syspar \in \Pareto} \hat \benefit(\syspar),
$
where $\Pareto$ is the set of all Pareto-optimal solutions. 

\subsubsection{Limitations in the Face of Externalities}
Consider $\syspar^{\circ}$, the system obtained when using the ``god's view'' values of social
utility:
$\syspar^{\circ} \define \arg \max_{\theta \in \Theta^\circ} \benefit(\syspar),$
with $\Theta^\circ$ being the set of the corresponding Pareto-optimal solutions
when using $\benefit$ in \Eqref{eq:benevolent-opt}.

Let $\Delta \benefit \define \benefit(\syspar^*) - \benefit(\syspar^\circ)$. We say that a system
with parameters $\syspar$ induces \newterm{externalities} on its environment when the social utility
of $\syspar$ is not equal to that of the system $\syspar^{\circ}$: $\Delta \benefit \neq 0$.

This definition is analogous to the neoclassical economic interpretation of externalities: due to an
``inefficiency,'' a partial optimum ($\syspar^*$) is achieved, that is different from the optimum if
no inefficiencies were present ($\syspar^{\circ}$)~\cite{Dahlman79}.  One can also parallel the
divergence between social-utility values in $\Delta \benefit$ to the Pigouvian ``divergence
of social and private costs''~\cite{Pigou20}.
Because we are analyzing harms, we focus on negative externalities: $\Delta \benefit < 0$.

\begin{figure}
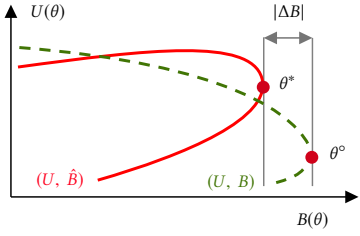

    \include{images/utilities}
    \caption{Pareto frontiers with real $(\sysutil, \benefit)$, and provider's version $(\sysutil,
    \hat \benefit)$. Misspecification of $\benefit$ results in externalities $\Delta B$,
difference in values of the benefit function between provider's system $\benefit(\syspar^*)$ and
$\benefit(\syspar^\circ)$.}
    \label{fig:utilities}
\end{figure}

Clearly, $\Delta \benefit = 0$ when $\benefit = \hat \benefit$. If $\benefit$ does not precisely
match $\hat \benefit$, however, there will likely be externalities present, as illustrated in
\Figref{fig:utilities}. In general, it is not known how exactly $\Delta \benefit$ is impacted by 
deviations of $\hat \benefit$ from $\benefit$. In Appendix~\ref{appx:sensitivity}, for a class of
strictly concave utility functions we show that the \newterm{sensitivity} of $\Delta \benefit$ to
infinitisemal error of $\hat \benefit$ is approximately quadratic in the magnitude of the error.
Intuitively, $\Delta \benefit$ grows quadratically fast as $\hat \benefit$ diverges.

\parait{Incomplete Information of Social Utilities.}
We first study the case in which the provider's view
of the social utility, $\hat \benefit$, does not fully reflect
the requirements and the context of the application domain.

Social benefit is often modeled as a function incorporating individual models of social utility. A
common assumption in neoclassical economics is that social utility is the sum
of individual utilities~\cite{Yew83}: $\hat
\benefit(\syspar) = \sum_{i \in \population} \hat \benefit_i(\syspar),$ where $\hat
\benefit_\misspecuser: \Theta \rightarrow \sR$ is a utility of individual $\misspecuser \in
\population$. In practice, however, the fair-by-design provider has \newterm{incomplete
information}: they are aware of the users, yet lack the full knowledge of their needs (similar to
``imperfect knowledge'' in economics and game theory~\cite{Phlips88}).
That is, they could misspecify $\hat \benefit_\misspecuser(\syspar)
\neq \benefit_\misspecuser(\syspar)$ for some user $\misspecuser \in \population$, inadvertently
ignoring the needs and well-being of this individual.
Consider a hypothetical fair-by-design predictive policing~\cite{LumW16} application that assumes
that $\hat \benefit$ is maximized when there is equality of false positives in patrol dispatches
across regions.  The resulting dispatching rates might be fair for this definition, but the overall
social benefit $\benefit$ might be unchanged as minorities could still be over-policed in terms
of the number of dispatches.

\parait{Omitting Impact on Non-Users and Environmental Impact.}
Another case is when the fair-by-design provider has structural lack of knowledge of the application
domain, e.g., knows the utility of their users, but not the utility of anything else
(i.e., of phenomena not shared with the machine). Thus, a fair solution
for the users could harm non-users as their benefits were never specified in the model $\hat \benefit(\syspar)$.
This is exemplified in the case of self-regulating housing markets, which could hypothetically be
optimized for fairness in acceptance rates for guests~\cite{airbnb-discriminates}, but disrupt
neighborhoods impacting housing conditions, especially for people with low incomes.

\subsubsection{Limitations Under Complete Knowledge}
Even in the presence of complete information regarding $\sysutil$ and $\benefit$, there can exist
externalities that the provider cannot mitigate.

First, the provider can maximize $\benefit$ and yet cause externalities if the system's goal
\emph{itself} is harmful~\cite{KeyesHD19}. For example, a facial recognition surveillance system
might be completely fair with respect to skin color, but it still causes risks and harms to the
population as a whole in terms of privacy loss.

Second, the fair-by-design model has as premise that there exists a solution $\syspar^{\circ}$ that, if not
maximizes, then at least satisfies minimum standards of everyone's benefit. However, this may not be
the case. Recall the examples about unsafe housing or bad jobs in
\Secref{sec:jackson-discrimination}. In those cases, the fair outcome might still be harmful
for all users.

\subsection{Limitations on Ideals}

Mitigating externalities becomes a greater issue when incentives, capacities, and power structures
are not aligned. One way this manifests itself is in the modeling of social benefit as a function.
Indeed, a function cannot encode all the nuance regarding human needs. Moreover, neither the provider nor
the users' utility functions model the political context or the power asymmetries in the environment
of the machine~\cite{power2004social}. These can heavily influence or skew what we assume to be the
ground truth $\benefit$. Thus, power and politics may come to render the system specification unfair
even when it was designed considering the perfect benefit function for users.

Furthermore, the fairness-by-design approach inherently assumes that the pro\-vi\-der always has
enough resources to implement the fair solution that maximizes social utility. However, it is
unreasonable to believe that such an assumption will hold in practice. For instance, even if Waze
was cognizant of the needs of all individuals and all the peculiarities in their streets,
it is unlikely that they could afford an operations in which all of those constraints are taken into
account or be capable of mitigating the impact of externalities due to the interaction of multiple
traffic beating applications.

Finally, not all service providers have the incentives to implement fair-by-design solutions. The
provider \emph{might not be incentivized} to care about social utility or could benefit from
externalizing certain costs.  In such cases, the magnitude of the externalities $\Delta \benefit$ is
likely to be more pronounced than in the case of incomplete knowledge.

%% file: images/utilities.tex
\tikzset{every picture/.style={line width=0.75pt}} 

\scalebox{0.75}{\begin{tikzpicture}[x=0.75pt,y=0.75pt,yscale=-1,xscale=1]

\draw [color={rgb, 255:red, 0; green, 0; blue, 0 }  ,draw opacity=0.5 ][line width=0.75]    (342,92.6) -- (342,192.73) ;

\draw [color={rgb, 255:red, 0; green, 0; blue, 0 }  ,draw opacity=0.5 ][line width=0.75]    (374.75,93) -- (374.75,192.73) ;

\draw [color={rgb, 255:red, 255; green, 0; blue, 0 }  ,draw opacity=1 ][line width=1.5]    (176.75,112.6) .. controls (264.75,100.6) and (332.67,92.8) .. (340.67,119.8) .. controls (348.67,146.8) and (289.38,171.73) .. (230.38,188.73) ;

\draw [color={rgb, 255:red, 65; green, 117; blue, 5 }  ,draw opacity=1 ][line width=1.5]  [dash pattern={on 5.63pt off 4.5pt}]  (177.67,99.47) .. controls (316.38,110.4) and (365.38,143.73) .. (374.33,170) .. controls (370.38,183.4) and (364.38,186.4) .. (350.38,190.4) ;

\draw  [draw opacity=0][fill={rgb, 255:red, 208; green, 2; blue, 27 }  ,fill opacity=1 ] (370,173.33) .. controls (370,170.94) and (371.94,169) .. (374.33,169) .. controls (376.73,169) and (378.67,170.94) .. (378.67,173.33) .. controls (378.67,175.73) and (376.73,177.67) .. (374.33,177.67) .. controls (371.94,177.67) and (370,175.73) .. (370,173.33) -- cycle ;
\draw  [draw opacity=0][fill={rgb, 255:red, 208; green, 2; blue, 27 }  ,fill opacity=1 ] (337.33,126.13) .. controls (337.33,123.74) and (339.27,121.8) .. (341.67,121.8) .. controls (344.06,121.8) and (346,123.74) .. (346,126.13) .. controls (346,128.53) and (344.06,130.47) .. (341.67,130.47) .. controls (339.27,130.47) and (337.33,128.53) .. (337.33,126.13) -- cycle ;
\draw [color={rgb, 255:red, 117; green, 117; blue, 117 }  ,draw opacity=1 ]   (345,93) -- (371.75,93) ;
\draw [shift={(374.75,93)}, rotate = 180] [fill={rgb, 255:red, 117; green, 117; blue, 117 }  ,fill opacity=1 ][line width=0.08]  [draw opacity=0] (8.93,-4.29) -- (0,0) -- (8.93,4.29) -- cycle    ;
\draw [shift={(342,93)}, rotate = 0] [fill={rgb, 255:red, 117; green, 117; blue, 117 }  ,fill opacity=1 ][line width=0.08]  [draw opacity=0] (8.93,-4.29) -- (0,0) -- (8.93,4.29) -- cycle    ;
\draw    (172,200) -- (402.75,200) ;
\draw [shift={(405.75,200)}, rotate = 180] [fill={rgb, 255:red, 0; green, 0; blue, 0 }  ][line width=0.08]  [draw opacity=0] (8.93,-4.29) -- (0,0) -- (8.93,4.29) -- cycle    ;

\draw    (172,200) -- (172,73.6) ;
\draw [shift={(172,70.6)}, rotate = 450] [fill={rgb, 255:red, 0; green, 0; blue, 0 }  ][line width=0.08]  [draw opacity=0] (8.93,-4.29) -- (0,0) -- (8.93,4.29) -- cycle    ;

\draw (358,123) node   {$\theta ^{*}$};
\draw (392,170) node   {$\theta ^{\circ }$};
\draw (359,76) node   {$|\Delta B|$};
\draw (375,216) node   {$B( \theta )$};
\draw (197,76) node   {$U( \theta )$};
\draw (205,188) node [color={rgb, 255:red, 255; green, 0; blue, 31 }  ,opacity=1 ] [align=left] {$\displaystyle ( U,\ \hat{B})$};
\draw (320,190) node [color={rgb, 255:red, 65; green, 117; blue, 5 }  ,opacity=1 ] [align=left] {$\displaystyle ( U,\ B)$};

\end{tikzpicture}}
\vspace{-2em}

%% file: parts/pots.tex
\section{Protective Optimization Technologies}
\label{sec:pots}
The systems view on algorithms in previous sections enables us to systematically explore a problem
space that had not been formalized before. A major source of problems is the use of optimization
techniques that help to capture and manipulate phenomena in the application domains for the \emph{extraction of value}.
This practice causes intentional or unforeseen changes to the environment which result in (often neglected) harms
to the environment.

We showed that existing fairness frameworks produce solutions that
can only address a subset of these harms, at the discretion of a service provider.
These frameworks have limited capability to mitigate
harms arising from inequities inherent to the application domain and from harmful impacts
introduced by the machine. By focusing solely on actions that can be taken by service providers, 
who have opposite incentives, 
fairness frameworks narrow both politics and contestation to the re-design of the algorithm which may not
always be the site of either the problem or the solution.

In this section, we discuss means to address these issues. Although these means could be of
socio-legal nature, we focus on technological approaches that we call \newterm{Protective
Optimization Technologies (POTs)}. Their goal is to shape the application domain in order to
reconfigure the phenomena shared with the machine to address harms in the environment.
POTs are designed to be deployed by actors affected by the optimization system.
As these actors directly experience the externalities, a) they have intimate knowledge of the
system's negative effects, b) they are in position to have a better view of their social utility
than a system provider can model---because it is their own utility~\cite{Dolan98}. Lastly, c) POTs do not rely on the incentives of the provider.

POTs seek to equip individuals and collectives with tools to counter or contest the externalities of the system.
Their goal is not to maximize benefits for both users and service providers, nor to find the best strategy to
enable optimization systems to extract value at minimum damage. POTs are intended to
\emph{eliminate} the harms induced by the optimization system, or at least \emph{mitigate} them. In
other cases, POTs may \emph{shift} the harms to another part of the problem
space where the harm can be less damaging for the environment or can be easier to deal with.
Finally, when service providers react to reduce the effectiveness of POTs, this very action
\emph{exposes} the service providers' need to maintain the power relationship and their capability
to manipulate the environment to their own benefit.

\begin{figure}
\centering
    \includegraphics[width=\columnwidth]{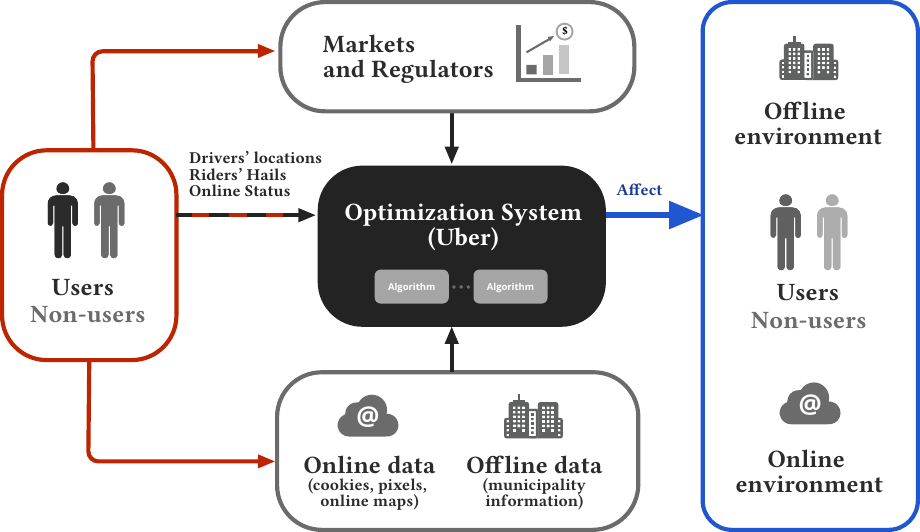}
    \caption{Uber: a complex optimization system. Inputs to Uber optimization (black), effects on
    the environment (blue), and alternatives to deploy POTs (red).}
    \label{fig:uber}
\end{figure}

\parait{POTs and Optimization Systems.}
To extract value through optimization, service providers obtain inputs from their environments that help them make
decisions. We consider three kinds of inputs: i) inputs that users generate when interacting with
the system, ii) inputs about individuals and environments received from third
parties~\cite{BarrenecheWilken15}, and iii) inputs from regulations and markets that define the
political and economic context in which the system operates. Note that third parties can be public
or private, and the data they provide can be gathered online or offline. We use Uber ride-hailing service~\cite{uber_url} as an example to showcase this complexity (\Figref{fig:uber}).

In order to maximize its profit, Uber optimizes the prices offered to riders and
the wages offered to drivers. Uber uses the following inputs (black arrows).  First, it
uses the direct inputs it receives from both riders and drivers.  Second, it uses data from other
sources such as online service providers, e.g., Google for maps and data that they might collect using
cookies or pixels on other sites their users visit~\cite{uber-cookies}. Uber also receives offline data from
parties like municipalities interested in promoting the use of Uber to reduce
costs of public transport~\cite{uber-innisifil}. Lastly, Uber uses inputs from the market and
regulators to evaluate the economic context in order to adjust wages and ride prices.

Uber ultimately uses these inputs and all the political and economic context in a combination of
managerial and mathematical optimization to deliver outcomes to the environment: match riders to
drivers and set ride prices. Reports and studies demonstrate that these outcomes cause
externalities: Uber's activity increases congestion and misuse of bike
lanes~\cite{uber-congestion}, increases pollution, and decreases public support for
transit~\cite{GrahelerME19,uber-bad}.


In the previous sections' vocabulary, Uber and its optimization techniques
are the machine; and the application domain comprises Uber drivers and riders, non-users, the
online and offline environments, and the market and regulatory frameworks in which Uber operates.
Roughly speaking, Uber's requirements $R$ are to match drivers to riders with associated dynamic pricing.
The specification $S$ defines what Uber applications should do to fulfill $R$.

Uber is known to have unfairness problems. For example, Rosenblat et al. show that customer-based
reviews are biased against minority drivers. As getting blocked from the system depends on these
reviews, even if the Uber's algorithms do not discriminate drivers on their attributes per se, the
rules of the system result in disparate impact on blocked drivers~\cite{RosenblatLBH17}.  Further,
even when Uber algorithms are fair, e.g., they reward all drivers the same irrespective of their
protected attributes, the optimization processes underlying Uber's operation result in unjust
outcomes: low wages~\cite{uber-wage}. The former is an externality stemming from structural biases in the application domain (thus a
hypothetical $S_\fair$ does not result on $R_\fair$); whereas the latter is a problem of incentives
misalignment.

In the Uber scenario, POTs can be deployed by users and non-users with the goal of changing the
phenomena captured by Uber. These can come in three forms (red lines in ~\Figref{fig:uber}): by
changing the inputs of the users to the system (e.g. the surge-induction POT~\cite{Uber} as we
describe shortly in \Secref{sec:wildpots}), by changing the online or offline signals gathered by Uber (e.g.,
mayors changing the city urban planning), or by affecting the market (e.g., by changing regulations
or mandating salary increases~\cite{RosenblatLBH17}).

\subsection{Examples of POTs}
\label{sec:wildpots}

\input{parts/wildpots}

Our vision for POTs systematizes the use of technologies as tools to both explore the
effects that algorithms and optimization systems have on our society, and the design of
countermeasures to contest their negative effects in the absence of regulatory or other
accountability measures.

We now revisit recent academic technologies, artistic interventions, and deployed tools,
that can be reframed as POTs (Table \ref{tab:wildpots} provides a summary). POTs formalize such technologies and interventions,
enabling the systematic study and design of such solutions. We illustrate this
in \Secref{sec:casestudies}, where we design two POTs from scratch.

These technologies have different origins. First, we observe that there are technologies proposed in
the academia that can be repurposed as POTs. For instance, in the field of computer security, research
that aims to protect against attackers gaming the YouTube algorithm~\cite{Fang18} can be repurposed
by users to fight against filter bubbles; in the field of adversarial machine learning, tools
developed to evade copyright detection~\cite{Saadatpanah19}, originally developed to strengthen DRM,
can be reframed as a way to prevent fair-use takedowns~\cite{fairuse_EFF}. Second, we draw from works produced by artists looking into the impact of technology on society. For
instance, counter-surveillance fashion that tricks facial recognition technologies~\cite{dazzle} can be
repurposed to evade discriminatory facial recognition algorithms or discriminatory uses of facial
recognition. Finally, we look to deployed technologies that are already countering optimization, either
intentionally or as a side effect. For instance, Jobscan~\cite{jobscan} assists job applicants in
getting past the automated sorting implemented by large companies and job-posting sites. This tool
could be repurposed to reduce the gender or racial bias reported for these tools~\cite{amazon}.

These examples can also be categorized based on the means used
to design the POT: those based relying on adversarial machine
learning~\cite{Sharif16,fu18, Fang18,Nogueira18,hyperface,jobscan}, 
and those that use heuristics to exploit the target optimization process. For instance, as a response to
low wages, Uber drivers have developed heuristics for inducing surges~\cite{Uber}.

Finally, some technologies can be deployed individually and others require collective action.
AdNauseam~\cite{adnauseam}, for example, reduces the
utility of ad networks by clicking all ads served to the user, flooding the network with
false information. Without a critical mass of users, however, AdNauseam's effect would not be
observable.

%% file: parts/wildpots.tex
\begin{table*}[h!]
\caption{Repurposing technologies as POTs. We detail the origin of the technology; the externality it addresses, the optimization system causing the externality and the desired outcome of the POT; deployment requirements (individual or cooperative action); and the underlying design techniques (counter-optimization or heuristics to decide how to shape the environment). }\label{tab:wildpots}
\resizebox{\textwidth}{!}{%
\begin{tabular}{lllllll}

 \textbf{Origin} & \textbf{Optimization System}                          & \textbf{Externality}                                            & \textbf{POT}                                                                                                                                                                 & \textbf{Desired Outcome}                                                                                                                                     & \textbf{Deployment}      & \textbf{Technique}     \\\hline
Academic & Face Recognition                              & Privacy, discrimination                     & Wear printed eyeglasses~\cite{Sharif16}                                                                                                     & Evade face detection                                                                                           & Individual & Optimization \\
Academic & Copyright Infringement Detection              & Fair use takedowns                                     & Adversarial examples~\cite{Saadatpanah19}                                                                                   & Avoid a fair use takedown                                                                                 & Individual & Optimization \\
Academic & Psychometric Profiling                        & Privacy, manipulations                          & Text style transfer~\cite{fu18, Nogueira18}                                                                          & Prevent from attribute inference                                                                                                                    & Individual & Optimization \\
Academic & YouTube Recommendations                       & Manipulation                                           & Poisoning~\cite{Fang18}                                                                                             & Breaking out of content bubbles                                                                                                 & Individual & Optimization  \\
Academic & Waze Routing                                  & Local traffic congestion                               & Sybil devices simulate traffic~\cite{Wang18}                                                                                     & Prevent routing into towns                                                                                   & Individual & Heuristic \\
Academic & GRE Scorer                                    & Biased grading system                                  & Generate essay to pass GRE~\cite{GRE} & Higher test score                                                                                                                                   & Individual & Heuristic     \\
Deployed & Ad Network                                    & Privacy, manipulations                          &Click on all ads~\cite{adnauseam}                                                                  & Ad Network destroyed                                                                                                                                & Collective & Heuristic     \\
Deployed & Uber Pricing System                           & Low wages                              & Shut off app, turn it back on~\cite{Uber}                                                                                                                           & Induce surge                                                                                                                                        & Collective & Heuristic    \\
Deployed & Instacart Pricing                             & Low wages      & Tip 22\cent~in app, cash at door~\cite{Instacart}                                                                                    & Fair pay for jobs                                                                                                                                & Collective & Heuristic     \\
Deployed & Automated Hiring                              & Bias, discrimination                       & Edit resume~\cite{jobscan}                                                                    & Flip automated hiring decision                                                                                       & Individual & Optimization \\
Deployed & Pokemon Go Resource Spawn                     & Unfairness                                             & Edit Open Street Maps~\cite{pogo}                                                                                                                                               & Encourage resources to spawn                                                                                                                        & Individual & Heuristic     \\
Deployed & FitBit for Insurance Premium                  & Privacy, surveillance                           & Spoof device location~\cite{fitbit}                                                                   & Get insurance benefits                                                                                                                & Individual & Heuristic    \\
Deployed & Pharma Optimizing Patents & End of humanity & Find potential drugs using ML~\cite{farm} &
Get drugs in the public domain & Individual & Mixed \\
Deployed & Insurance Coverage Optimization & High costs of treatment & Doctors changing claim codes~\cite{WyniaCVW00} & Get higher reimbursements & Individual & Heuristic \\
Artistic & Face Recognition                              & Privacy, surveillance                           & Scarf that is classified as a face~\cite{hyperface}                                            & Evade face detection                                                                                                                             & Individual & Optimization  \\
Artistic & Face Recognition                              & Privacy, surveillance                           & Camouflage to cover features~\cite{dazzle}                                                                & Evade face detection                                                                                                                             & Individual & Heuristic     \\
Artistic & Autonomous Cars                               & Exploration risks                                      & Ground markings~\cite{bridle}           & Trap autonomous cars                                                                                                                             & Individual & Heuristic   \\
\end{tabular}}
\end{table*}

%% file: parts/eval.tex
\section{Designing Protective Optimization Technologies: Case studies}
\label{sec:casestudies}

In this section, we show how, given an optimization system and a negative
externality, one can design new POTs from scratch.

Like many of the technologies and interventions in Table~\ref{tab:wildpots}, we make use of
optimization techniques to design the operation of POTs. Starting from the model introduced in
~\Secref{sec:fairness-critique}, we model the optimization system's objective as a function
$\sysutil(\sysinputvec; \syspar): \sysinputspace^m \times \Theta \rightarrow \sR,$ where
$\sysinputspace^m$ denotes the space of environment inputs to the system (i.e., phenomena that are
sensed by the machine), and the vector $\sysinputvec \in \sysinputspace^m$ is a set of inputs
coming to the system. Each concrete input $\sysinput_i$ can come from a different source (users,
non-users, or other actors in the environment). For simplicity, we model the time dimension through
discrete time steps, and we assume that there are two possible time steps: $0$ and $1$.

To maximize its gain, the optimization system strives to solve a mathematical optimization problem
$\syspar^*_t = \arg \max_{\syspar} \sysutil(\sysinputvec_t; \syspar)$ for a set of inputs
$\sysinputvec_t$ at each point in time $t$.
Given this system, the goal of a POT is finding actionable---feasible and
inexpensive---modifications to the inputs of the optimization system so as to maximize social utility
that we denote as $\benefit_\pot(\syspar)$. Note that this definition of social utility needs
not to correspond to the social utility considered by the optimization system ($\hat
\benefit(\syspar)$ in Section~\ref{sec:fairness-critique}). 
We consider that each input has an associated modification cost $\modcost(\sysinput_i \rightarrow
\sysinput_i'): \sysinputspace \times \sysinputspace \rightarrow \sR^+$, that represents how hard
it is to modify it. The cost of changing a set of inputs $\modcost(\sysinputvec \rightarrow
\sysinputvec')$ can be any function of the individual costs.

In this model, we define the POT design as a bi-level multi-objective optimization problem:
 \begin{equation}
     \begin{aligned}
         & \min_{\sysinputvec'}
         \{ \modcost(\sysinputvec \rightarrow \sysinputvec'), -\benefit_\pot(\theta^*_{t+1}) \} \\
         \text{ s.t. } & \theta^*_{t+1} = \arg \max_{\syspar_{t+1}} \sysutil(\sysinputvec',
         \syspar_{t+1})
     \end{aligned}
     \label{eq:minmax}
 \end{equation}
where $\sysinputvec$ is the vector of inputs if no intervention happened, and $\sysinputvec'$ are
possible vectors of modified inputs. Parameter $\syspar_{t+1}$ represents the system's state in the
next step, after the POT has been deployed through the modified inputs $\sysinputvec'$.

We instantiate this problem for two different use cases, one where the POT can be deployed by an
individual, and one where effective deployment requires a collective. The code for both case studies
is available online.\footnote{\url{https://github.com/spring-epfl/pots}}

\input{parts/eval-waze.tex}
\input{parts/eval-loan.tex}

%% file: parts/eval-waze.tex
\subsection{Thwarting Traffic from Routing Apps}
\label{sec:eval-waze}
In our first case study, we look at Waze, a crowdsourced mobile routing application that optimizes the
routes of its users based on information about traffic, road closures,
etc.~\cite{waze_url}. Waze causes negative externalities for residents of towns and neighbourhoods
that are adjacent to busy routes. For example, take the town of Leonia, New Jersey, USA, which lies just outside
one of the bridges into New York City.
As Waze rose in popularity and directed an increasing amount of users through the town when the
highway was busy, the town became crowded during rush hours.  To prevent Waze traffic, the
town was briefly closed off to non-local traffic, which was determined illegal~\cite{Shkolnikova18}.
In this section we propose a solution for discouraging Waze from selecting routes through the
town while minimizing the impact on its inhabitants.

\parait{Problem Setup.}
We set up this problem as a planning problem in which the town's traffic network is modeled as a weighted directed graph, and the goal is to increase the cost of paths between the highway ramps.  We define Waze's utility $\sysutil$ as the capability to provide fastest routes for its users. Routing through town can increase this
utility when it takes less time than traveling via the highway.
The POT is designed to increase the minimum time through town so that Waze users are not routed
through it. We define the social utility $\benefit_\pot$ as a binary variable that takes value 1 when no vehicle is routed through the town (i.e., the cost in time of traversing the town is greater than traversing the highway) and 0 otherwise.

Let $\graph = (\vertices, \edges, \timecost)$ be a weighted directed graph representing the traffic
network within the town. Each edge $(x, y) \in \edges \subset \vertices \times \vertices$ represents a
road segment, and each vertex represents a junction of segments.  The edges have associated time to
pass the segment given by the function $\timefunc(x, y)$.  We define the
\newterm{time cost} of traversing a path in the graph as the total time to pass its edges:
$\timecost(\graphpath) \define \sum_{(x, y) \in \graphpath} \timefunc(x, y)$.

Let $\source, \sink \in \vertices$ be the source and sink vertices that represent the entry to the
town from the highway $(\source)$, and the exit to return to the highway $(\sink)$.
Let the time to travel from point $\source$ to $\sink$ via the highway be $\timethreshold$.
While we do not know the routing algorithm used by Waze, we assume that Waze will send users through the town when the path through town is quicker than the highway. That is, if there is a path $\graphpath$ from $\source$ to $\sink$ in $\graph$ with cost $\timecost(\graphpath) < \timethreshold$, Waze routes users through the town.

\subsubsection{Avoiding Routing via Planning}
We aim to transform the graph $\graph$ into a graph $\graph'$ such that the time cost of any path
$\graphpath'$ from $\source$ to $\sink$ in $\graph'$ is $\timecost(\graphpath') \geq \timethreshold$.
We focus on what the town can control: time to traverse a road segment.  We express these changes as the increase in time that it takes to traverse
a segment, $\timedelta(x,y)$. We abstract the
exact method for adding time to the segment, which could be changes to speed limits, traffic lights,
stop signs, speed bumps etc. We construct $\graph'$ by modifying the time values in the original
graph $\graph$: $\timefunc'(x, y) = \timefunc(x, y) + \timedelta(x, y),$ where $\timefunc'$ is a
function representing the edge time in $\graph'$.

We acknowledge that some roads are more costly to change than others by associating a modification
cost to every road. In practice, this cost will be specified by the town. We use \emph{length} of a
road segment as such a cost, capturing that changing longer roads will likely have more impact on
the town.  Let $\interdictvar(x, y) \in \{0, 1\}$ be binary \newterm{interdiction variables} that
represent whether we are modifying the edge $(x, y)$ in graph $\graph'$. To express the cost
$\modcost$ of modifying a graph $\graph$ into $\graph'$ we use the following modification-cost
function: $\sum_{(x, y) \in \edges} \edgecost(x, y) \cdot \interdictvar(x, y),$ where $\edgecost:
\edges \rightarrow \sR^+$ represents the cost of modifying the edge $(x, y)$.

We now formalize the POT through posing the multi-objective optimization problem in
\Eqref{eq:minmax} in a constraint form: minimize $\modcost$ subject to the $\benefit_\pot$
constraint, given that Waze is maximizing $\sysutil$:
\begin{equation}\label{eq:orig-waze-problem-speed}
    \begin{aligned}
        & \min_{\interdictvar(\cdot) \in \{0, 1\}} \sum_{(x, y) \in \edges} \edgecost(x, y) \cdot \interdictvar(x, y) \\
        \text{ s.t. } & \timefunc(\graphpath) \geq \timethreshold
        \text{ for any path $\graphpath$ from $\source$ to $\sink$}
    \end{aligned}
\end{equation}

This is equivalent to the problem known as \newterm{shortest-path network
interdiction}~\cite{Golden78, FulkersonH77, IsraeliW02, WeiZXYZ18}. In the form equivalent to ours it
can be solved as a mixed-integer linear program (MILP)~\cite{IsraeliW02, WeiZXYZ18}. We refer to
Appendix~\ref{appx:waze} for the exact formulation of the MILP in our context.  We use
ortools~\cite{ortools_url} for specifying this MILP in Python, and the CBC open-source
solver~\cite{CBC} for solving.

\subsubsection{Empirical Evaluation}
We apply this POT to three towns, all of which reported issues with Waze: Leonia, NJ, USA;
Lieusaint, France; and Fremont, CA, USA.  We retrieve the map data for each via the Open Street Maps
API~\cite{osm_url}.

To assign $\timefunc(x, y)$ to each segment $(x, y)$, we estimate the time it takes to traverse the
segment using its length and type (e.g., we consider residential streets to have a speed of
25mph) and add extra time for traversing the intersection.  We infer the intersection time using the
travel time from Google Maps for accuracy.  For each city, we run the solver for different $\timethreshold$
values, corresponding to different travel times on the highway. Larger values of $\timethreshold$
require more changes to roads in the town.

\begin{figure}[h!]
\centering
\begin{subfigure}{.4\columnwidth}
  \centering
  \includegraphics[height=\textwidth]{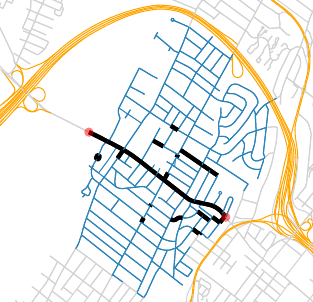}~%
 \label{fig:sub1}
\end{subfigure}%
\hfill
\begin{subfigure}{.4\columnwidth}
  \centering
  \includegraphics[height=\textwidth]{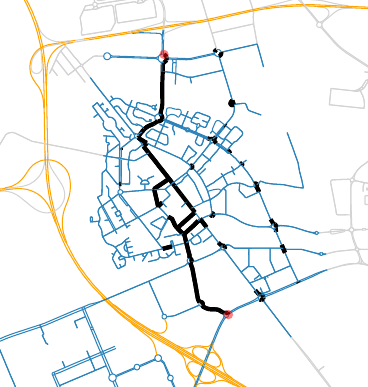}~%
  \label{fig:sub2}
\end{subfigure}
\caption{Solutions for Leonia, NJ (left) and Lieusaint, France (right), when the time of road
    segments is allowed to be increased by 75\%. Town streets are marked in blue, highways in
    orange, and surroundings in grey. The red dots signify $a$ and $b$, the closest points to the
    highway in the town.  The roads marked by thick, black lines are the optimal set of segments in
    which the time should be increased.}
\label{fig:frnj}
\end{figure}

For simplicity of implementation, we choose two points at the edge of the town (red dots in
\Figref{fig:frnj}) and not the sink/source points, which would be on the highway, for the solver.
We then add the time that it takes to travel from the actual source point on highway to the first
point and from the second point back to the sink point in order the calculate the results.  For
Leonia, for instance, we approximated this as 30 seconds on each side of the town.

We consider scenarios in which the town is able to increase the time that it takes to traverse each
road segment by 25\%, 50\%, and 75\%.  For each town we found the value of $\timethreshold$ for
which Waze begins sending cars through the town and the value of $\timethreshold$ in which no
further changes to the road can prevent Waze from sending its users through the town.

The graph for Leonia, NJ contains 251 vertices and 661 edges. Given the parameters we choose, without
any changes to the town, Waze begins to send its users through Leonia when $\timethreshold=4.0$.
That is, it normally takes 4 minutes to travel through Leonia, so when it takes longer than 4
minutes to traverse the highway, Waze routes users through town. This corresponds to traffic
traveling at about 31mph on the highway.  If we limit the amount of time that can be added to
traverse a road segment to 25\% of the original time, we can prevent cars from being routed through
Leonia until the average highway speed has reached 26mph ($\timethreshold=5.1$). That is, we can
find a solution that prevents routing through town as long as the highway speed is greater than
26mph. For 75\% time increase, we can change the town such that traffic will not be routed through
Leonia until the average speed on the highway has fallen to 19mph. This solution for Leonia is shown
in Figure \ref{fig:frnj}.

Lieusaint is larger than Leonia, with 645 vertices and 1,248 edges. Given the parameters we
choose, Waze routes its users through Lieusaint when  $\timethreshold=7.0$, which corresponds to the
speed on the highway dropping to 16mph. Allowing the road segments to be lowered by 75\%, we
can prevent traffic from being routed through Lieusaint until the highway speed has dropped below
10mph ($\timethreshold=12.0$) (See Lieusaint in Figure \ref{fig:frnj}). We report the solution for Fremont, CA, a significantly larger town, in Appendix~\ref{appx:waze}.

\begin{figure}[h!]
\centering
   \includegraphics[width=.85\columnwidth]{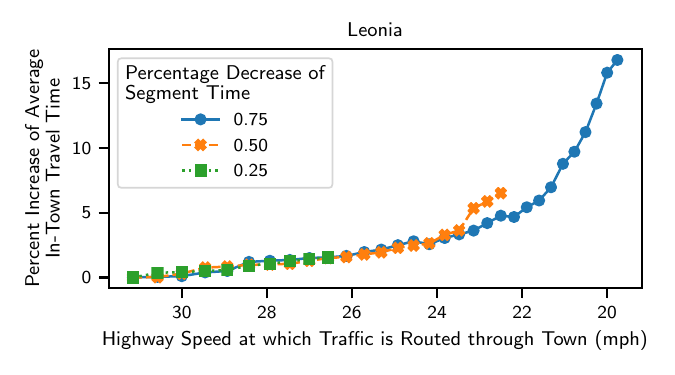}
   \includegraphics[width=.85\columnwidth]{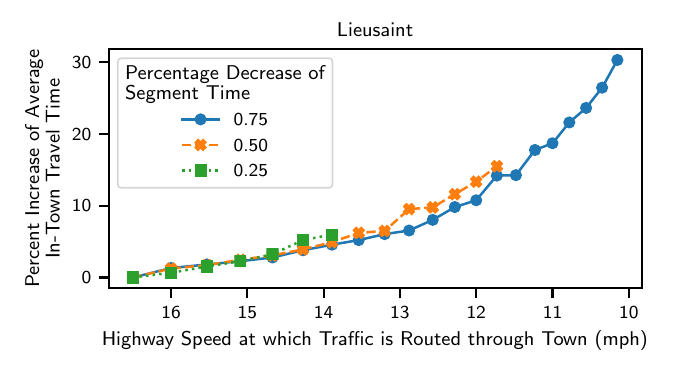}
\caption{Effect of Changes on In-Town Travel.}
\label{fig:cost_dist}
\end{figure}

Finally, we measure the cost of implementing these solutions (\Figref{fig:cost_dist}).  For each
town, we consider the impact to the town to be how much longer, on average, it takes to travel
between any two points in town. We compute the shortest path between every pair of points in the
town and average these times before and after the POT solution. We then compute the percentage
increase from the initial average time to the post-POT average time. The higher impact solutions are
those which will tolerate a lower highway speed. That is, they will prevent cars from being routed
through the town at lower highway speeds. We see that even though allowing road segments to take 75\% longer to traverse can prevent cars from entering the town at a lower highway speed, the impact to the town is much higher. The inhabitants of Lieusaint also suffer more from the changes than the residents of the much smaller Leonia.

\subsubsection{POT Impact and Limitations}
The intervention we propose in this section focuses on alleviating the problems routing applications
cause in \emph{one} town. The POT would support this town's government to decide which changes
should be implemented in the city layout and traffic rules so that the cost of traversing the city
becomes undesirable for passthrough drivers. While this intervention indeed mitigates the effect of
external traffic on the target town, it is likely that then vehicles are routed elsewhere, i.e., the
POT \emph{shifts} the harms from this particular town to other regions. Moreover, we acknowledge
that our POT can only help in cases when vehicles can still circulate on the highway.
The moment the congestion forces vehicles to halt, the town becomes the better option regardless of any modification on its road network.

%% file: parts/eval-loan.tex

\subsection{Reducing False Negatives in Credit Scoring}
In this case study, we explore solutions for countering harmful effects of a machine-learning
\newterm{credit scoring} system: A system that a bank uses to decide how to price a loan application
according to the predicted risk.  The underlying algorithms that support such decisions are designed
to maximize banks' profits and minimize their risks.  These algorithms can be
discriminatory~\cite{CitronP14}, or cause feedback loops for populations disadvantaged by the
financial system~\cite{Poon16}. These harms are often caused by inputs that represent unjust
realities which are propagated to the model's decisions.

\parait{Problem Setup.}
We model the credit scoring system as a classifier $f_\syspar(x)$ that takes as input the
information about the loan applicant and the loan details, $x$, and outputs a confidence score for
whether the applicant would repay the loan or not. This function optimizes the bank's utility that
we model here as the negative empirical loss over historical data:
$
  \theta^* \define \arg \min_{\theta \in \Theta} \sum_{(x, y) \in X} L(x, y; \theta),
$
where $\theta$ is the parameters of the classifier, $L$ is the loss function, and $X$ is the
bank's dataset of historical loans ($x$) and their repayment or default outcomes ($y$).
We assume that the classifier is retrained as new data points arrive.

We model the harms of the system as the rate of false negatives (wrong default
predictions) for economically disadvantaged populations. In this POT, we counter this problem using
adversarial machine learning techniques deployed by a collective of individuals $\collective$ with
the means to take and repay loans, as explained shortly. This POT aims to increase the social
utility defined as the negative loss on a \newterm{target group} $G$: $\benefit_\pot(\syspar) \define
-\sum_{(x, y) \in G} L(x, y; \syspar),$ where $G$ is the disadvantaged subset of applicants (in this
case study, we define disadvantaged as having little funds in the bank account) who were wrongfully
denied a loan.  This POT can be thought as promoting equality of false-negative rates between the
target group and everyone else~\cite{HardtPS16, Chouldechova17}, with the difference that we do not
limit our view of externalities to a commonly protected subgroup, i.e., economical disadvantage is
not commonly considered as protected.

\subsubsection{Reducing False Negatives with Adversarial Machine Learning}
We first identify what inputs ($\sysinputvec$ in the abstract model) can be modified by the
collective $\collective$ deploying the POT. First, the deployers can only add new inputs to the
dataset by taking and repaying loans. Second, the demographic attributes of these added loan
applications have to be similar to those of individuals in $G$. Thus, the POT must inform
the collective $\collective$ about who and for which loans they should apply for and repay, in such
a way that they reduce the false-negative rate on the target group after retraining.  This POT is
idealistic in that it assumes that this collective will include applicants of diverse backgrounds to
be able to provide different inputs to the classifier.  However, in the absence of other means of
feedback, communication, and accountability, it represents the only means to influence an unjust
system. It is also consistent with the existing practices people resort to in this
setting~\cite{Friedman17, OShea18}.

Finding which inputs to inject into a training dataset to modify its outputs is known as
\newterm{poisoning} in adversarial machine learning~\cite{PapernotMSW16}.  Typically, poisoning
attacks aim to increase the average error of the classifier or increase the error on specific
inputs~\cite{BiggioR18}.  We note that our use of poisoning is different.  First, we poison to
\emph{decrease the error} for a given target group. Second, our use of poisoning is not adversarial.
On the contrary, we use it to protect users against harmful effects of the model.

With this in mind, we design the POT using the following bi-level optimization problem:
\[
    \begin{aligned}
        & \min \objective(\syspar^*) = \sum_{(x, y) \in G} L(x, y; \theta^*) + \lambda \reg(\syspar^*) \\
        \text{ s.t. } & \theta^* = \arg \min_{\theta \in \Theta} \sum_{(x, y) \in X \cup \poisonset} L(x, y; \theta) \\
                      & f_{\theta'}(x) = \text{`accept'} \text{ for all $(x, y) \in \poisonset$} \\
                      & \poisonset \subset \pool, |\poisonset| \leq n
    \end{aligned}
\]
where $\theta'$ is the current parameters of the classifier, $\poisonset$ is the \newterm{set of
poisoned applications}, $n$ is the maximum number of poisoned applications, and $\pool$ is a
\newterm{set of feasible loan applications}. That is, we minimize the classifier's loss for the
target group, where the classifier $\theta^*$ is trained using the poisoned examples $\poisonset$.
In our evaluation below, we additionally make use of a regularizer $\reg$ to minimize the effect of poisoning
on any other applications: $\reg(\syspar) \define -\sum_{x, y \in X} L(x, f(x); \syspar).$

This formulation makes two assumptions. First, we assume that when designing the POT we have access
to the training dataset of the provider $X$, and we can obtain retrained loan-approval models.  We
consider this assumption reasonable as poisoning attacks tend to transfer even if the dataset and
model do not match the real ones, especially when the models have low
complexity~\cite{DemontisMPJBONR19}.  Second, the added loan applications must be feasible: there
has to exist a person in the $\collective$ with demographics required for this loan application.

We solve this problem by scoring each example in
$\pool$ according to the value of our optimization objective $\objective(\syspar^*)$, retraining
$\syspar^*$ for each example, and then employing a greedy algorithm to assemble $\poisonset$ that
satisfies the constraints. We refer to the Appendix~\ref{appx:loans} for the details of the algorithm.

\subsubsection{Empirical Evaluation}
We create a simulated loan approval system using the German credit risk dataset from the UCI Machine
Learning Repository~\cite{uciml}. This dataset contains 1,000 feature vectors representing loan
applications, including applicants' job type, bank account details, gender, etc., and
loan details: amount, duration, and purpose.  Each has a binary label encoding whether the
loan was repaid (70\%) or not (30\%).  We implement the loan approval system as a
logistic regression classifier.  For our simulation, we split the dataset into the bank's training
dataset $X$ (800 examples), and the test set (200).  The classifier achieves 75.5\% test accuracy.
We obtained the best results with the use of $\lambda = 0.5$ as the trade-off parameter for the
regularization term (out of 0.25, 0.5, 0.1) in $\objective(\syspar^*)$.

We simulate the set of feasible applications $\pool$ using a subset of \emph{successfully repaid}
applications from users that are not in the target group by generating all possible changes of
modifiable attributes.

We evaluate our POT in two settings: a ``clean'' setting in which the only applications received by
the bank are those from the collective ($X \cup \poisonset$); and a ``noisy'' setting in which other
people take loans between the time when the POT is deployed and the time when the classifier is
retrained. We report the results in the noisy setting for 10 and 50 additional loans (1.25\% and
6.25\% of the original dataset, respectively).  We repeat the noisy experiments 10 times in which we
draw the ``noise'' loan applications at random.

\begin{figure}
  \centering
  \includegraphics[width=.75\linewidth]{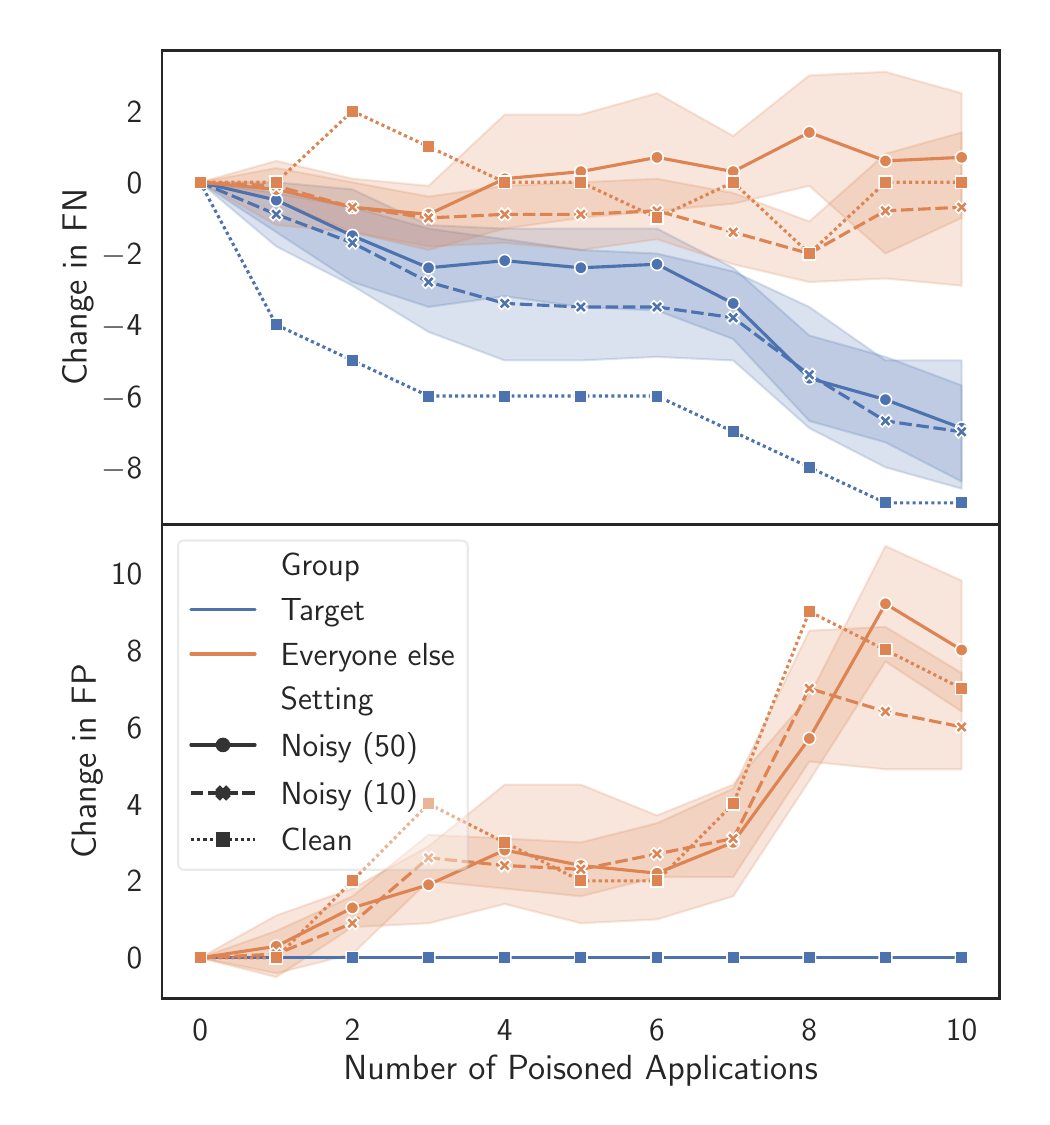}~
  \caption{Effects of the POT on the error rates of the classifier for the target group. Top: decrease
    in the number of false negatives (wrongly denied loans). Bottom: increase in the number of
    false positives (wrongly granted loans).}
  \label{fig:poisoning}
\end{figure}

We present the effect of poisoning on the target group and everyone else in \Figref{fig:poisoning},
top. Unsurprisingly, the more poisoning applications there are, the more significant the effect.
With 10 poisoning applications (1.25\% of the training dataset), our algorithm reduces the number of
false negatives in the target group by 9; in the presence of noise this decreases to 7. The false
negatives in the rest of the dataset are on average not impacted by our POT.

\subsubsection{POT Impact and Limitations}
Unfortunately, the POT increases the false positives (\Figref{fig:poisoning}, bottom). That is, it
\emph{shifts} the harm to the bank, which would give more loans to people who cannot repay,
increasing its risk.  This effect sharply increases with the seventh application (out of 10). This
is due to the poisoning inputs starting to change the model parameter corresponding to the loan
purpose. In turn, this increase in false positives could lead to adverse social impact over time if
banks try to counter the POT effect~\cite{LiuDRSH19}. The POT deployers could adjust the trade-off
parameter $\lambda$ to control for the side-effects.

To be consistent with the fairness literature~\cite{HardtPS16, YangS17}, we assume that the target
group is interested in getting higher credit ratings (through decrease in false-negatives). This
model that postulates the access to credit as beneficial, however, is na\"ive.  Even though the
loan-making process that relies on risk-based pricing has economic backing, by definition it implies
that the less finanically advantaged people will get more expensive loans.  Hence, an intervention
that aims at increasing inclusion of disadvantaged populations in the lending process can be seen as
an instance of \newterm{predatory inclusion}~\cite{Poon16, SeamsterCC17, Taylor19}. Even if it results in
lower loan prices in the short term, it can lead to dispossession in the long run. When harms are viewed through this lens, It is not clear
if any technological intervention is capable of counteracting such systems.

%% file: parts/discussion.tex
\label{sec:discussion}

Our inspiration to provide a more holistic view on optimization systems and their harms comes from
works that point to the logic and potential impact of optimization systems. In particular,
Poon has drawn attention to the ways in which optimization systems are driven by outcomes, as
exemplified in our utility functions in \Secref{sec:fairness-critique}. This allows for techniques like operational control and
statistical management to be the primary mode with which machines interact with phenomena in the
world~\cite{Poon16}. As a result, these techniques function both as a means for engineering and as ``a
mathematical state that poses as a solution to political contention''~\cite{mckelvey2018internet}.
Optimization is a technique long established in, for example, resource allocation. However, its increasing supremacy in systems that reconfigure every aspect of life
(education, care, health, love, work etc.) for extraction of value is new and refashions all social, political, and
governance questions into economic ones. This shift allows companies to commodify aspects of
life in a way that conflates hard questions around resource allocation with maximization of profit
and management of risk~\cite{gandy2010engaging}. The impact is subtle but fundamental, as evident in
the way even we start framing complex and historical questions of justice in terms of utility
functions.

To ground ourselves in the world, we used the requirements engineering model of Michael A. Jackson,
aligning it with calls for decentering technology and centering communities in the design of
systems~\cite{pena2019decentering}. We extended the model to ensure that we are not suggesting that
problems and solutions are neatly separable and void of power and political economy~\cite{Powles18}.
However, many elements that might be crucial for conceptualizing optimization systems are still
missing. The ontology offers few concepts to capture matters around data, machine learning, or
services, and it does not provide deeper insights into addressing issues like
fairness or justice. It is, however, a nod to the importance of ontological work for systematizing
reflections on our frameworks~\cite{CorbettG18,GreenH18}.

To capture the political economy of optimization systems, we turned to utilitarian models and
calculations of externalities.
Such models are commonly used both in mathematical and managerial forms of optimization and are the
cornerstone of neoclassical economics. However, utilitarian models have been thoroughly critiqued
for, among others, perpetuating inequalities~\cite{power2004social}. Most prominently, Sen has highlighted the limitations of 
assessing value through consequences, 
assessing value through subjective utility, 
maximizing welfare without regard for its distribution, 
and fetishizing resources over relations between resources and people~\cite{sen79}.
Overall, utilitarian approaches are weak in capturing collective interests, social well-being, forms
of power, and subjugation. Given these critiques, the central role that these models play in
designing large-scale optimization systems is a problem in and of itself.


One possible way forward is to consider alternative economic models for the design and evaluation of
 systems, e.g.,~\cite{sen84,Diaz15,power2004social,fleurbaey2018social}. 
 POTs depend and build on the existence of such alternative economic models and the availability of
collectivity, altruism and reciprocity.  They assume there is something to be gained both
individually and collectively, dismissing the selfish agents presumed in utilitarian approaches. In
fact, we struggled to express POTs in the utilitarian logic: if we optimize for the utility of the
service provider it is hard to justify any POT that may reduce the utility of the
service provider.

Beyond economic gains, POTs strategically support people's agency. Optimization systems offer little agency to effectively contest their value proposition~\cite{Thomas19,Myanmar18,Miracle18,CitronP14} and offer more optimization as solutions for externalities~\cite{Vincent18}. POTs can be used to exercise some agency towards an unaccountable~\cite{oddNumbersPasquale} and ``authoritative system''~\cite{ziewitz2019rethinking}. Nevertheless, service providers may argue that POTs are gaming the system. 
Our focus is on social-justice contexts, in which POTs can be cast as ``weapons of the geek''
for the least equipped to deal with the consequences of optimization \cite{Coleman15}. POTs can also
serve to expose systems' injustices, achieving transparency and accountability goals. In that sense,
they also can come to act like \newterm{rhetorical software} ``that turn the logic of a system
against itself [...]  to demonstrate its pathology''~\cite{Tseng19}. This includes making apparent
the damaging results of utilitarian forms of governance prominent in optimization systems.

Despite their positive potential, designing and deploying POTs is not trivial. By virtue of
modifying, subverting, or sabotaging an optimization system, POTs may elicit transitions in the
system state that result in externalities. If several POTs are deployed and enter in an arms race,
those agents with the most knowledge and resources are likely to deploy the most aggressive and
effective POTs and have the most leverage. This undermines the ability of less powerful populations,
who may need POTs the most, to have any effect on the system.  This signals that well-thought POTs
must be built to provide less powerful actors with the means to respond to the potential abuse of
power by those that have more capabilities.

Just as much, the multi-input form of most optimization systems poses a serious challenge: when
optimization is based on continuous tracking across many channels, POTs cannot be built short of
creating ``optimized doubles'' of entities in the environments~\cite{ruckenstein2014visualized}. The
fact that a whole infrastructure for optimizing populations and environments is built in a
way that minimizes their ability to object to or contest the use of their inputs for optimization is
of great concern to the authors---and should be also to anyone who believes in choice,
subjectivity, and democratic forms of governance.

%% file: parts/appendices.tex
\numberwithin{equation}{section}

\section{Sensitivity to Misspecifications}
\label{appx:sensitivity}

We theoretically estimate the impact of misspecifications on the severity of externalities. For
that, we use influence functions from the toolkit of robust statistics~\cite{Huber11}.

We assume that the utility functions $\sysutil$, $\benefit$, and $\hat \benefit$ are strictly
concave and twice-differentiable; and we strengthen the ideal property of the fair-by-design
provider.  Besides picking the Pareto-optimal solution that maximizes their model of social utility,
we now assume the near-optimality of the social-utility objective for this solution: $\nabla \hat
\benefit(\syspar^*) \approx 0$. That is, the system $\syspar^*$ is close to the \newterm{ideal
solution} for the social-utility objective $\hat \benefit(\syspar)$~\cite{Miettinen12}.

Consider a \emph{partially corrected} optimization objective:
\[
    \max_{\syspar \in \Theta} \{ \sysutil(\syspar), \hat
    \benefit(\syspar) - \varepsilon \hat \benefit(\syspar) + \varepsilon \benefit(\syspar) \},
\]
where we move a pointmass $\varepsilon$ away from the provider's model of the social utility to its
``god's view'' value. Let us take its a Pareto-optimal solution $\syspar^*_\correction$ that has
highest benefit. We define the \newterm{influence function} of $\Delta \benefit$ as follows:
\[
    \influence \define \frac{d}{d\varepsilon} [\benefit(\syspar^*_\correction)
    - \benefit(\syspar^*)] \,.
\]
This models how fast the magnitude of the externality grows as more weight is
given to the corrected $\benefit$ in the optimization problem.

Let us restate a known property of Pareto-optimal solutions due to Kuhn and Tucker~\cite{KuhnT14}:
\begin{theorem}[\cite{KuhnT14}]\label{thm:kt}
    Let $\syspar^*$ be a Pareto-optimal solution to the optimization problem of the form:
    \[
        \max_{\syspar \in \Theta} \{ \sysutil(\syspar), \benefit(\syspar) \}
    \]

    Then there exists $\lambda \in [0, 1]$, such that the following holds:
    \[
        \lambda \nabla \sysutil(\syspar^*) + (1 - \lambda) \nabla \benefit(\syspar^*) = 0
    \]
\end{theorem}

Let us denote by $\hessian f(x)$ the Hessian matrix of $f$ at $x$, and for convenience set
$\hessian_{\syspar, \lambda} := \lambda \hessian \sysutil(\syspar) + (1 - \lambda) \hessian \hat
\benefit(\syspar)$.  Additionally, denote by $\Delta \syspar := \syspar^*_\correction - \syspar^*$
the difference in system parameters coming from the corrected and the original optimization
problems. We can now present our estimate for the influence function.

\begin{statement}
Using linearization techniques, we can obtain the following approximation for the influence function
for some $\lambda \in [0, 1]$:
\[
    \influence \approx -\nabla \benefit(\syspar^*)^\intercal
    [\hessian_{\syspar^*, \lambda}]^{-1} \nabla \benefit(\syspar^*)
\]
\end{statement}
\paragraph{Derivation}
Using \Thmref{thm:kt}, we can say there exists $\lambda \in [0, 1]$ such that:
\begin{equation}\label{eq:kt-cond-app}
    \lambda \nabla \sysutil(\syspar^*_\correction) + (1 -
    \lambda) \left(\nabla \hat \benefit(\syspar^*_\correction) - \varepsilon \nabla \hat
    \benefit(\syspar^*_\correction) +
    \varepsilon \nabla \benefit(\syspar^*_\correction)\right) = 0
\end{equation}

We now rewrite \Eqref{eq:kt-cond-app} in terms of the original system parameters $\syspar^*$ using a
first-order Taylor approximation:
\[
    \begin{aligned}
        0 & = \lambda \nabla \sysutil(\syspar^*) + (1 -
        \lambda) \left(\nabla \hat \benefit(\syspar^*) - \varepsilon \nabla \hat \benefit(\syspar^*) +
        \varepsilon \nabla \benefit(\syspar^*)\right) \\
          & + \left[\lambda \hessian \sysutil(\syspar^*) + (1 - \lambda)\left(
            \hessian \hat \benefit(\syspar^*) - \varepsilon \hessian \hat \benefit(\syspar^*) +
            \varepsilon \hessian \benefit(\syspar^*)
        \right)\right] \cdot \Delta \syspar
    \end{aligned}
\]

We can rearrange and further approximate following Koh and Liang~\cite{KohL17}, keeping in mind that
$\varepsilon$ is small:
\[
    \begin{aligned}
        \Delta \syspar & \approx -\varepsilon [\hessian_{\syspar^*, \lambda}]^{-1} \nabla
    [\benefit(\syspar^*) - \hat \benefit(\syspar^*)] \\
        & \approx -\varepsilon [\hessian_{\syspar^*, \lambda}]^{-1} \nabla
    \benefit(\syspar^*)
    \end{aligned}
\]

We now approximate the influence function using its first-order Taylor expansion and
the obtained expression for $\Delta \syspar$:
\[
  \begin{aligned}
    \influence =
    \frac{d}{d \varepsilon} [\benefit(\syspar^*_\correction) - \benefit(\syspar^*)]
    & \approx \frac{d}{d \varepsilon} \left[\benefit(\syspar^*) + \nabla \benefit(\syspar^*) \cdot
    \Delta \syspar \right] \\
    & = \nabla \benefit(\syspar^*) \cdot \frac{d}{d \varepsilon} \Delta \syspar \\
    & = - \nabla \benefit(\syspar^*)^\intercal [\hessian_{\syspar^*, \lambda}]^{-1} \nabla
        \benefit(\syspar^*)
  \end{aligned}
\]

\qed

\begin{statement}
    Given the assumptions on $\sysutil, \benefit, \hat \benefit$ and $\syspar^*$, our linear
    approximation for the influence function of $\Delta \benefit$ is asymptotically lower bounded as
    follows:
    \[
        \influence = \Omega (\parallel \nabla \benefit(\syspar^*) \parallel^2)
    \]
\end{statement}

\begin{proof}
As $\hessian_{\syspar^*, \lambda}$ is negative-definite by the concavity assumption and the fact
that convex combinations preserve concavity, so is its inverse. Hence, $-[\hessian_{\syspar^*,
\lambda}]^{-1}$ is positive-definite. By a lower bound of a symmetric positive-definite quadratic
form we have:
\[
  \begin{aligned}
    \frac{d}{d \varepsilon} \Delta_\benefit & \approx \nabla \benefit(\syspar^*)^\intercal
    [\hessian \objective(\syspar^*)]^{-1} \nabla \benefit(\syspar^*) \\
    & = \Omega(\parallel \nabla \benefit(\syspar^*) \parallel^2 )
  \end{aligned}
\]
\end{proof}

For concave functions, $\parallel \nabla \benefit(\theta^*) \parallel$ can serve as a measure of
error of the solution $\syspar^*$~\cite{BoydV04}, which confirms our intuition.

\section{Details for the Traffic Thwarting Case Study}
\label{appx:waze}

\subsection{MILP Formulation}
With a simple reparameterization, it is possible to formulate the optimization problem as follows:
\begin{equation}\label{eq:orig-waze-problem-time}
    \begin{aligned}
        & \min_{\interdictvar(\cdot) \in \{0, 1\}} \sum_{(x, y) \in \edges} \edgecost(x, y) \cdot \interdictvar(x, y)  \\
        \text{ s.t. } & \sum_{(x, y) \in \graphpath} [ \timefunc(x, y) + \interdictvar(x, y) \cdot \timedelta(x,
        y)] \geq \timethreshold \\
        & \text{ for any path $\graphpath$ from $\source$ to $\sink$},
    \end{aligned}
\end{equation}

In this form, the optimization problem is the shortest-path interdiction problem~\cite{Golden78,
FulkersonH77}, and can be solved as an MILP~\cite{IsraeliW02, WeiZXYZ18}:
\begin{equation}\label{eq:waze-problem-milp}
    \begin{aligned}
        & \min_{\interdictvar(\cdot) \in \{0, 1\}} \sum_{(x, y) \in \edges} \edgecost(x, y) \cdot \interdictvar(x, y)  \\
        \text{ s.t. } & \pathvar(t) - \pathvar(s) \geq \timethreshold \\
        & \pathvar(y) - \pathvar(x) \leq \timefunc(x,
            y) + \interdictvar(x, y) \cdot \timedelta(x, y) \quad \forall (x, y) \in \edges \\
        & \pathvar(v) \in \sR \quad \forall v \in \vertices,
    \end{aligned}
\end{equation}
where $\pathvar(v)$ are additional \newterm{vertex-potential variables} that represent the
smallest time cost for getting from $\source$ to $v$ in graph $\graph'$.

Assume that each edge $(x, y)$ is associated with a length defined by $\lengthfunc(x, y)$, and a
speed limit $\speedlimfunc(x, y)$. In the case of changing the speed limits through $\speeddelta$,
$\timedelta(x, y)$ can be obtained from $\lengthfunc$, $\speedlimfunc$ and $\speeddelta$ as follows:
\begin{equation}
    \timedelta(x, y) = \frac{\lengthfunc(x, y) \cdot \speeddelta(x, y)}{\speedlimfunc(x, y)^2 -
    \speedlimfunc(x, y) \cdot \speeddelta(x, y)}
\end{equation}

\subsection{Evaluation Details for Fremont, California}
\label{appx:cali}
\begin{figure*}[h!]
    \includegraphics[width=0.8\textwidth]{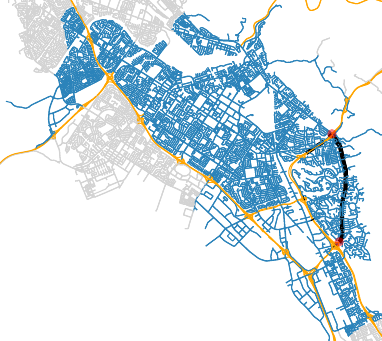}~%
    \caption{Solution for Fremont, California (in black)}\label{fig:ca}
\end{figure*}

The graph for Fremont, CA, USA, is much larger than for the other towns considered in our
evaluation, with a total of 9,215 nodes and 19,313 edges. The normal time from $\source$ to $\sink$
in the town is 8.5 minutes. Figure \ref{fig:ca} shows the optimal set of roads to lower the speed limit on for $\timethreshold=15.25$, using a 75\% decrease in time for the allowed road changes. 

\section{Details for the Credit-Scoring Case Study}
\label{appx:loans}
We detail the heuristic algorithm we use to solve the optimization problem of the POT in
Algorithm~\ref{alg:poisoning}. To compute the scores, we retrain a classifier for each example $(x,
y) \in \pool$. In our case of the logistic regression as the bank's model, retraining is
inexpensive. For more complex models, approximation techniques can be used~\cite{DemontisMPJBONR19}.

\begin{algorithm}
    \begin{enumerate}
        \item $S = \textsf{PriorityQueue}()$
        \item for $(x, y) \in \pool:$
        \item \quad continue if $f(x; \theta') \neq \text{`accept'}$
        \item \quad simulate $\theta^*_{(x, y)} = \arg \min \sum_{X \cup \{(x, y)\}} L(x', y'; \syspar)$
        \item \quad compute the score $\objective(\theta^*_{(x, y)})$
        \item \quad add $(x, y)$ to $S$ along with the computed score
        \item $\poisonset := \{\}$
        \item while $|\poisonset| \leq n$:
        \item \quad remove $(x^*, y^*)$ with the lowest score from $S$
        \item \quad add $(x^*, y^*)$ to R.
    \end{enumerate}
    \caption{Algorithm for selecting poisoning loan applications in order to reduce the
    false-negative rate on the target group.}
    \label{alg:poisoning}
\end{algorithm}